\documentclass{mem}
\usepackage{natbib}\usepackage{txfonts}\usepackage{balance}
\usepackage{graphicx}
\usepackage[a4paper,breaklinks,dvipdfm]{hyperref}
\idline{0}{00}
\begin{document}
\def\teff{$T\rm_{eff }$}
\def\kms{$\mathrm {km s}^{-1}$}

\title{
Extragalactic Globular Cluster Populations from High Resolution Integrated Light Spectra
}

   \subtitle{}

\author{
J. E.\,Colucci\inst{1}, 
R. A. \, Bernstein\inst{1},
A. \, McWilliam\inst{2},
\and J. G. \, Cohen\inst{3}
          }

  \offprints{J. E. Colucci}

\institute{
University of California, Santa Cruz Department of Astronomy/UC Observatories,
1156 High Street,
Santa Cruz, CA 95064, USA
\and
The Observatories of the Carnegie Institute of Washington, 813 Santa Barbara Street, Pasadena, CA 91101-1292, USA
\and
Palomar Observatory, Mail Stop 249-17, California Institute of Technology, Pasadena, CA 91125, USA. 
\email{jcolucci@ucolick.org}
}

\authorrunning{Colucci }

\titlerunning{GC Integrated Light Abundances}

\abstract{ We present a comparison of high-resolution,
  integrated-light, detailed chemical abundances for Galactic and
  extragalactic globular clusters in both massive galaxies and dwarf
  galaxies. We include measurements of Fe, Ca, Si, Na, and Al for
  globular cluster samples in 
  the Milky Way, M31, Large Magellanic Cloud, and NGC 5128.  These and
  other recent results from our group on M31 and NGC 5128 are the
  first chemical abundances derived from discrete absorption features
  in old stars beyond the Milky Way and its nearest neighbors.  These
  abundances can provide both galaxy enrichment histories and
  constraints on globular cluster formation and evolution.
  \keywords{Stars: abundances -- Galaxies: star clusters -- Galaxies:
    abundances -- Galaxies: formation } }
\maketitle{}

\section{Introduction}
 
Unresolved globular clusters (GCs) are luminous and therefore observationally accessible to large distances, unlike individual red giant branch stars.  This means that they are useful tools for  learning about the chemical enrichment and formation history of other galaxies when it is not possible to use high resolution spectra of individual stars.  Detailed abundances of key elements in old stars are in fact impossible to obtain in  any other massive galaxy beyond the Milky Way (MW), even in our nearest neighbor galaxy, M31.

With the development of our original technique for abundance analysis
of high resolution integrated light (IL) spectra of GCs, we can now
make significant advances in chemical evolution studies of distant
massive galaxies.  Our technique has been developed and demonstrated
on resolved GCs in the MW and Large Magellanic Cloud (LMC)
in a series of papers \citep{2,3,4,5,6, 7}.  These works show that the
IL analysis provides accurate Fe abundances and [X/Fe] ratios to
$\sim$0.1 dex for [Fe/H] of $-2$ to $+0$.  
This now makes it possible
to use diagnostic abundance ratios in old stellar clusters to gain
quantitative constraints on star formation histories in
galaxies up to 4 Mpc away.
Here we compare abundance results for key elements in 4 galaxies: the
Milky Way, M31, the LMC, and NGC 5128.

\section{Data \& Analysis}
The MW and LMC data presented here were obtained using the
echelle spectrograph on the duPont telescope at Las Campanas. These
data are described fully in \cite{3}, \cite{5}, and \cite{6}.  Our
sample of 27 GCs in M31 were observed with HIRES on the Keck I
telescope, and are described in \cite{4} and \cite{nic12}.  The first
4 GCs of our latest sample in NGC 5128 are also included here (Colucci et al. 2013, in prep) and
were obtained with the MIKE spectrograph on the Magellan Clay
Telescope in 2004-2005.  Exposure times in NGC 5128 were between
9$-$15 hours per GC, and the signal-to-noise ratio at 6000 \AA~
is $\sim$60-80.  Data were reduced with standard routines in
the MIKE Redux pipeline.
  
The methods we use in our IL abundance analysis are described in detail in
\cite{3,4,5,7}.  To summarize briefly, we measure
absorption line equivalent widths (EWs) using the
semi-automated program GETJOB \citep{8} and compare those with 
light-weighted, synthesized EWs that we calculate using 
an updated version of our ILABUNDS code \citep{3}, which calls 
synthesis routines from MOOG \citep{moog}.
We also synthesize spectral regions for direct comparison
to our spectra to assess blending and continuum placement.

\section{Abundances Across Galaxy Types} 
\subsection{M31 and the MW (Massive Spirals)}

We present abundances of Fe, Ca, and Si for 27 M31 GCs in the left
panels of Figure \ref{fig:abund}. Our IL abundances for MW GCs and
stellar MW abundances from the literature are also shown, for
comparison.  For the $\alpha$-elements Ca and Si, we find a plateau
value that is similar to the value in MW GCs and field stars from
\cite{venn} and \cite{pritzl}; this plateau is consistent with rapid
early star formation in M31.
There is also a clear knee visible in the M31 [Ca/Fe] values that
closely matches the plateau shown by MW field stars.

In Figure \ref{fig:abund}, we also show abundances for Na and Al.
Star-to-star variations in the light elements Na, O, Mg, and Al ---
specifically, depletion of Mg and O and enhancement of Na and Al ---
are a well known phenomenon in GCs  \citep[see][]{gratton}. In the IL abundances, these trends manifest as [Mg/Fe] being
lower than other [$\alpha$/Fe], and [Na/Fe] and [Al/Fe] higher than
expected compared to MW halo stars. We first published evidence of
this effect in the IL of GCs in \cite{4,7}.  Critically,
this implies that Mg is a {\it bad} proxy for [$\alpha$/Fe] in IL GC
spectra.  Preliminary results by \cite{larsen} recently confirm our
results.  Figure \ref{fig:abund} shows that the [Al/Fe] and [Na/Fe]
values for M31 GCs are also elevated, particularly Na at high [Fe/H].
We can therefore infer that most of the M31 GCs host star-to-star
abundance variations.

\vspace{-0.2cm}
\subsection{The LMC (Dwarf Irregular)}
\begin{figure*}
\includegraphics[clip=true,scale=0.38]{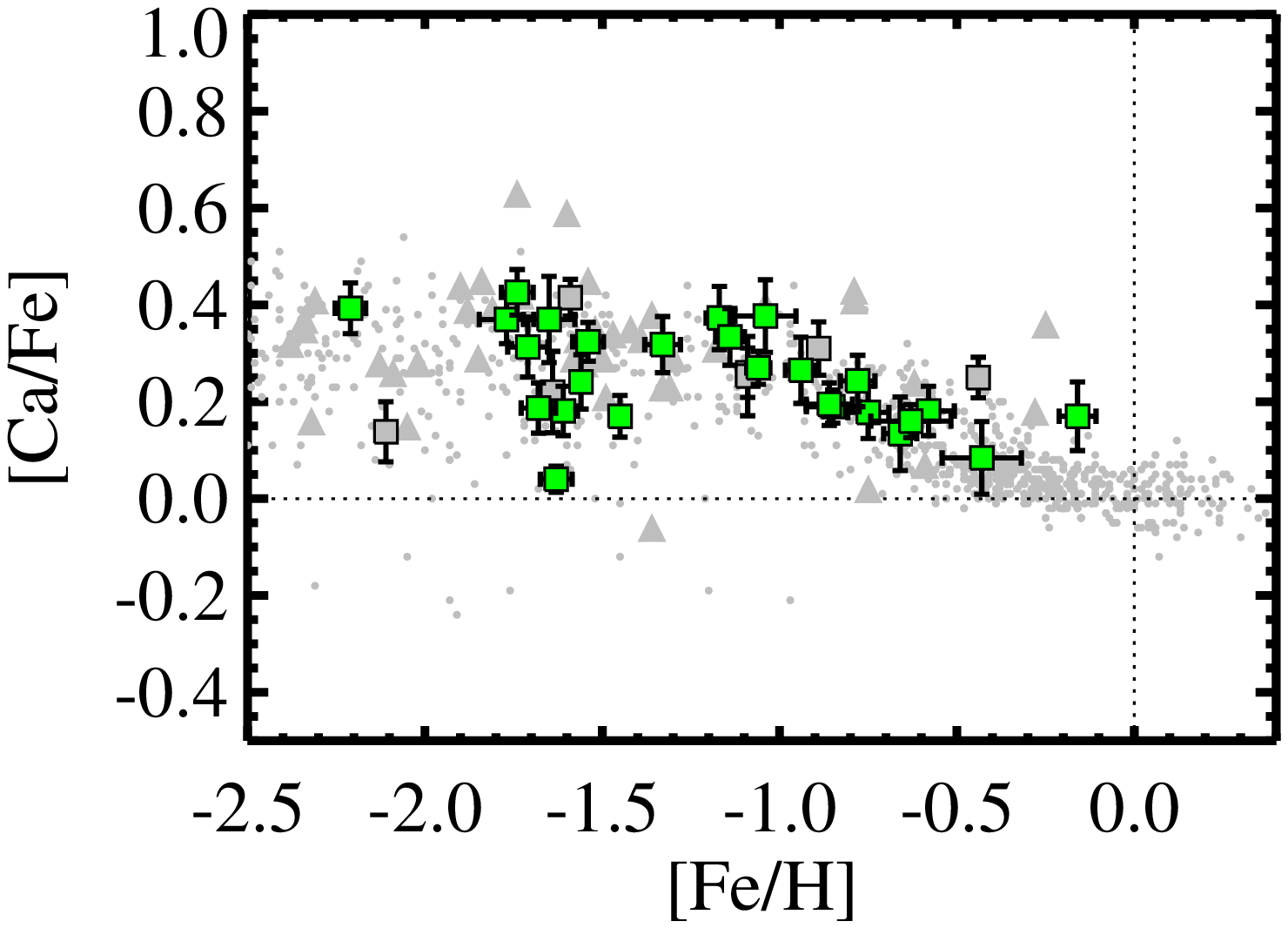}
\includegraphics[clip=true,scale=0.38]{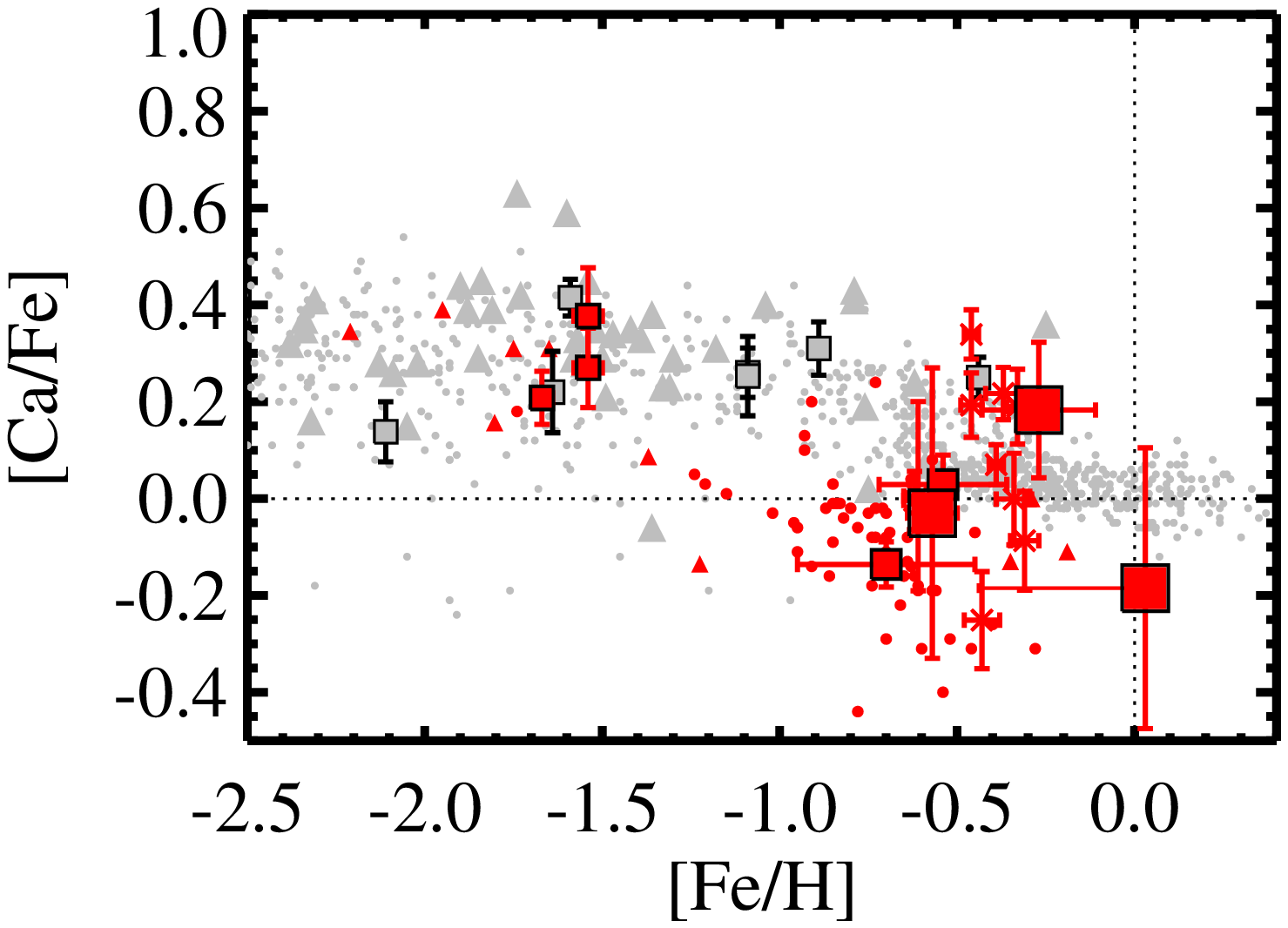}
\includegraphics[clip=true,scale=0.38]{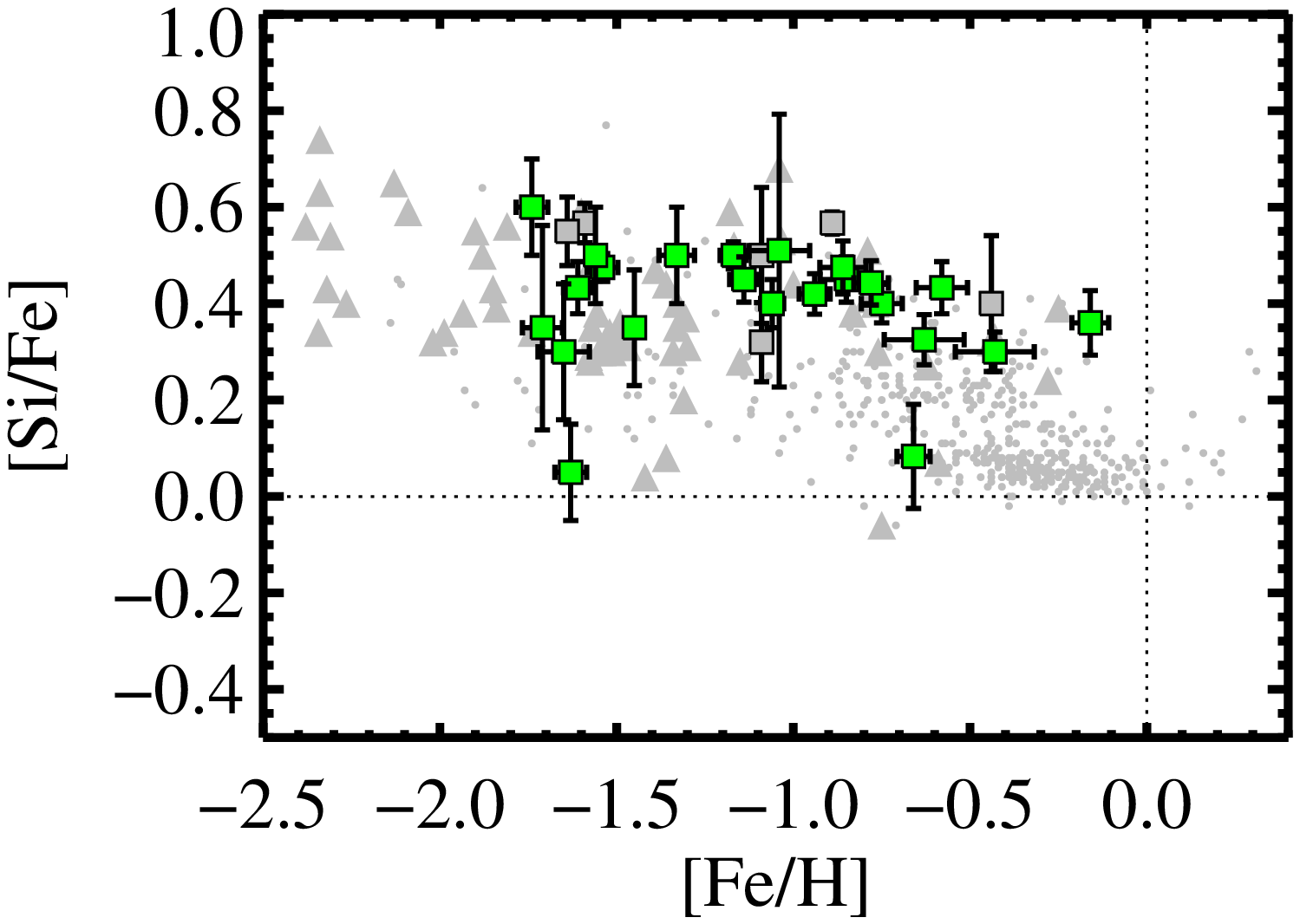}
\includegraphics[clip=true,scale=0.38]{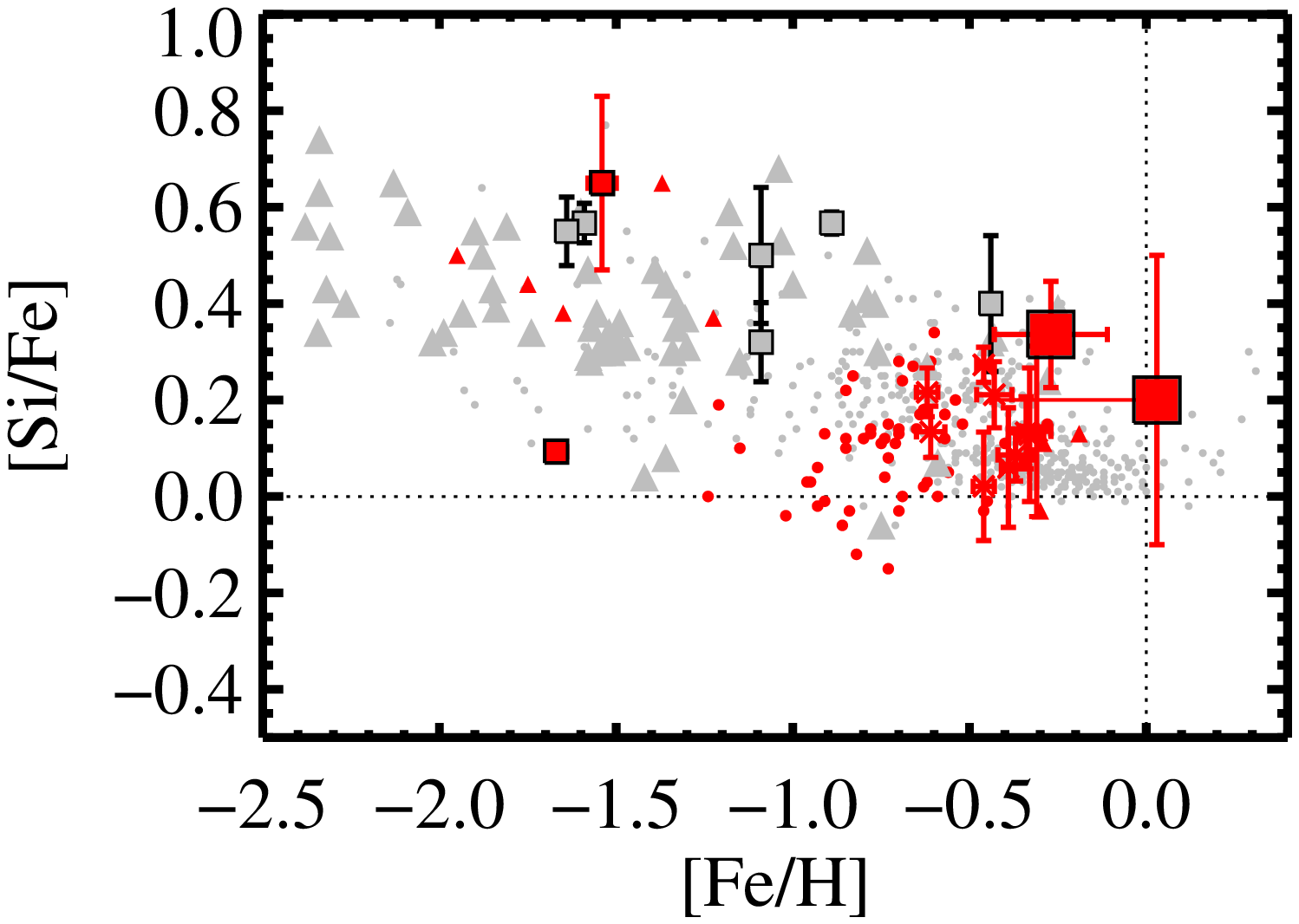}
\includegraphics[clip=true,scale=0.38]{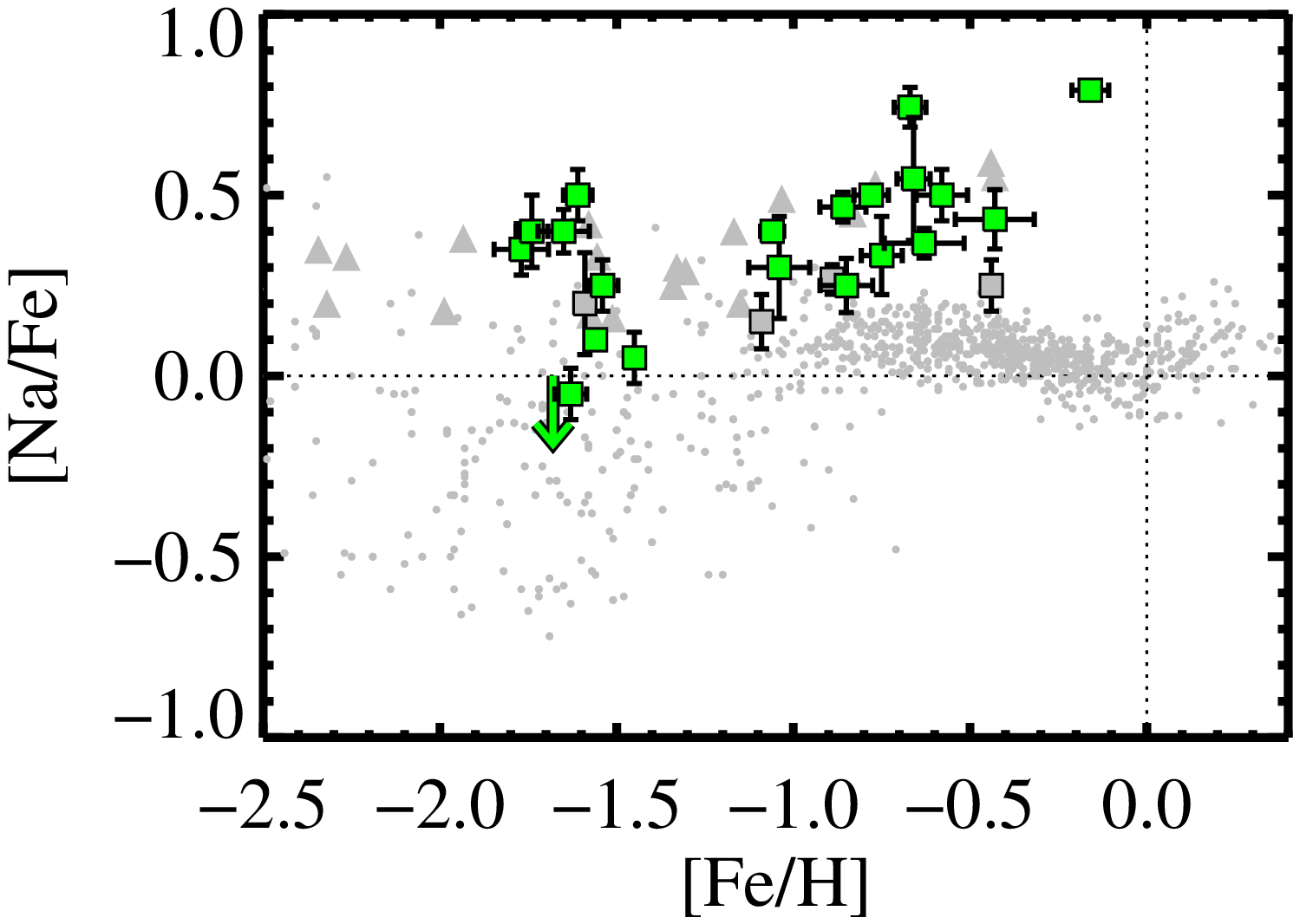}
\includegraphics[clip=true,scale=0.38]{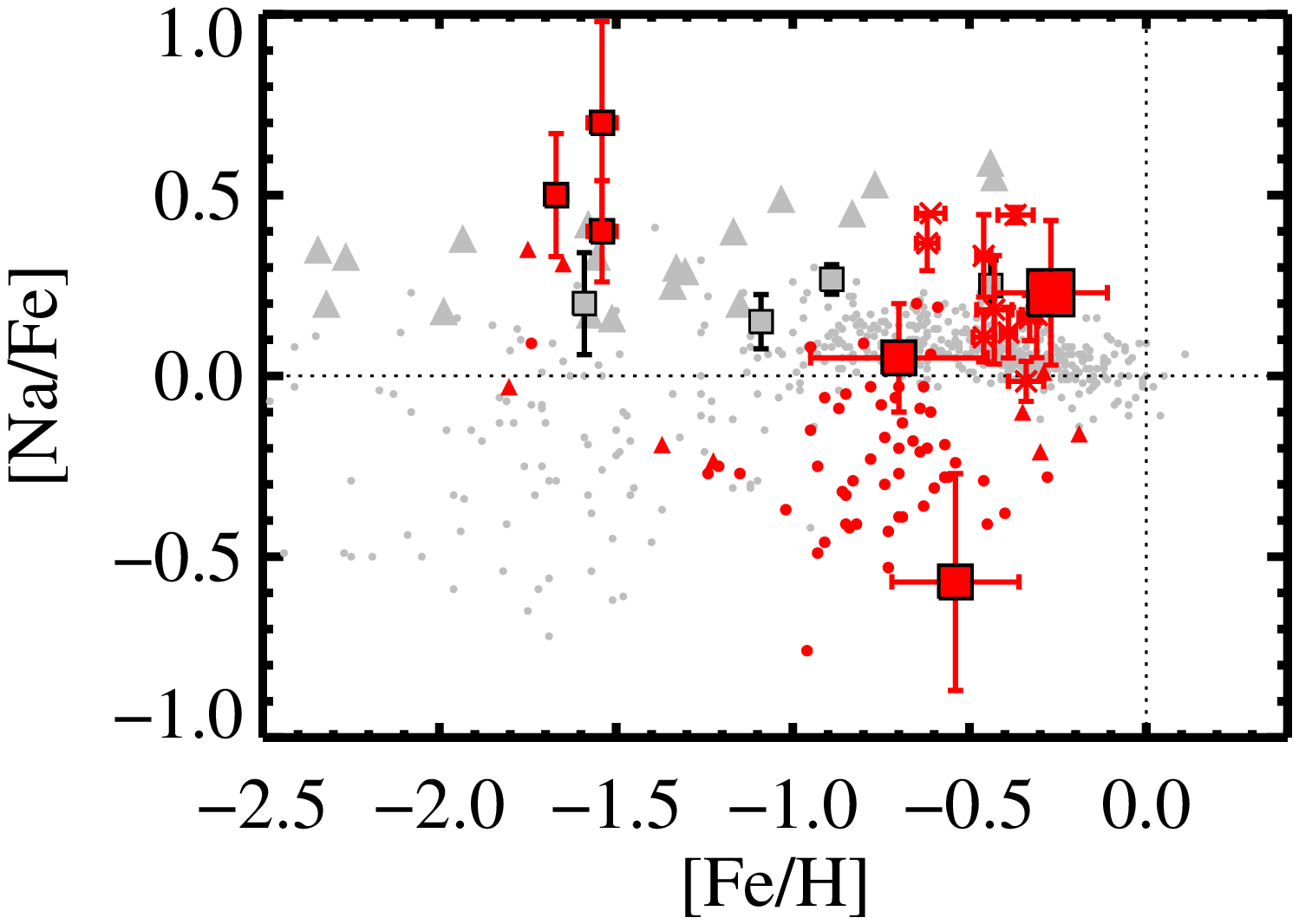}
\includegraphics[clip=true,scale=0.38]{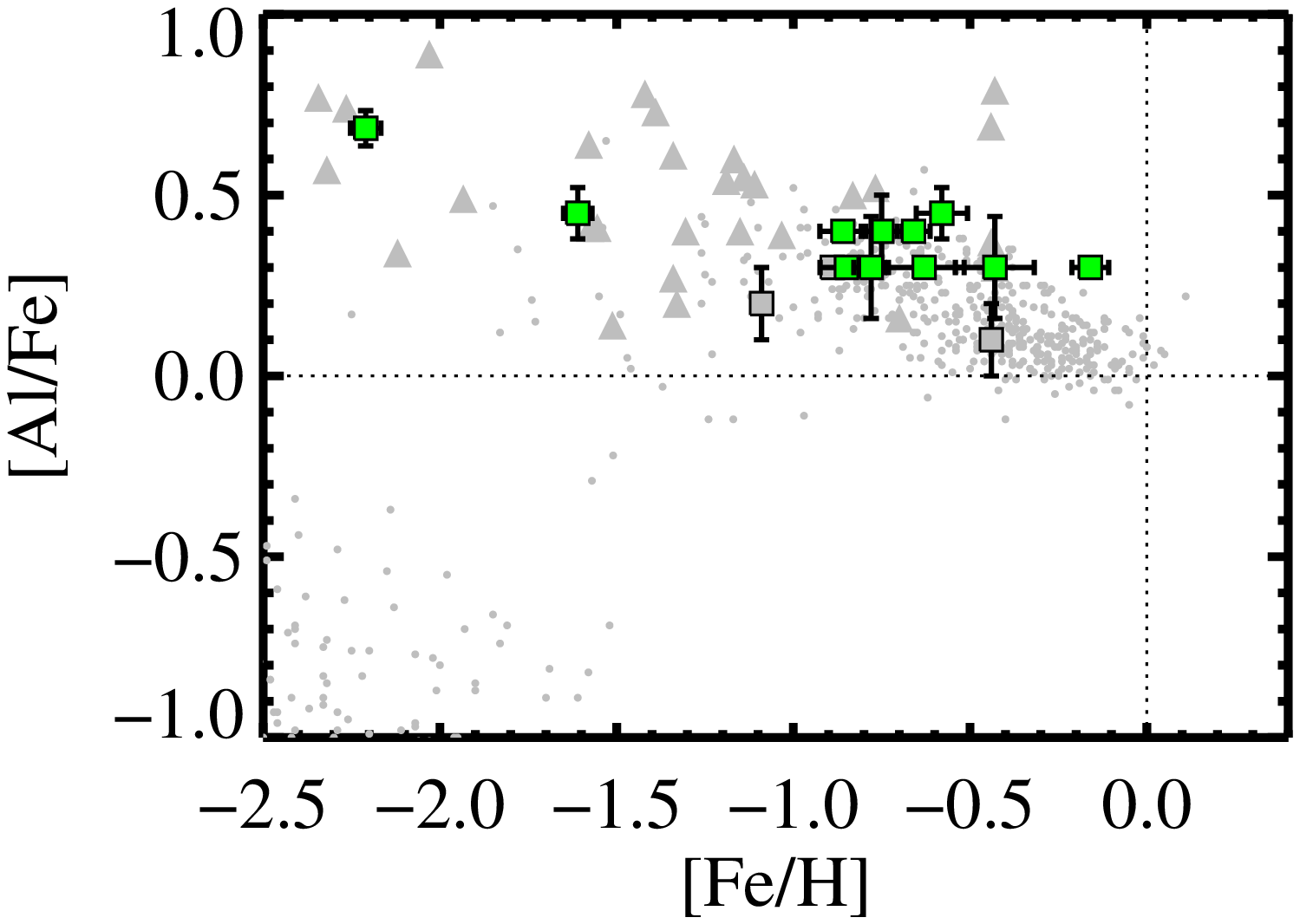}
\includegraphics[clip=true,scale=0.38]{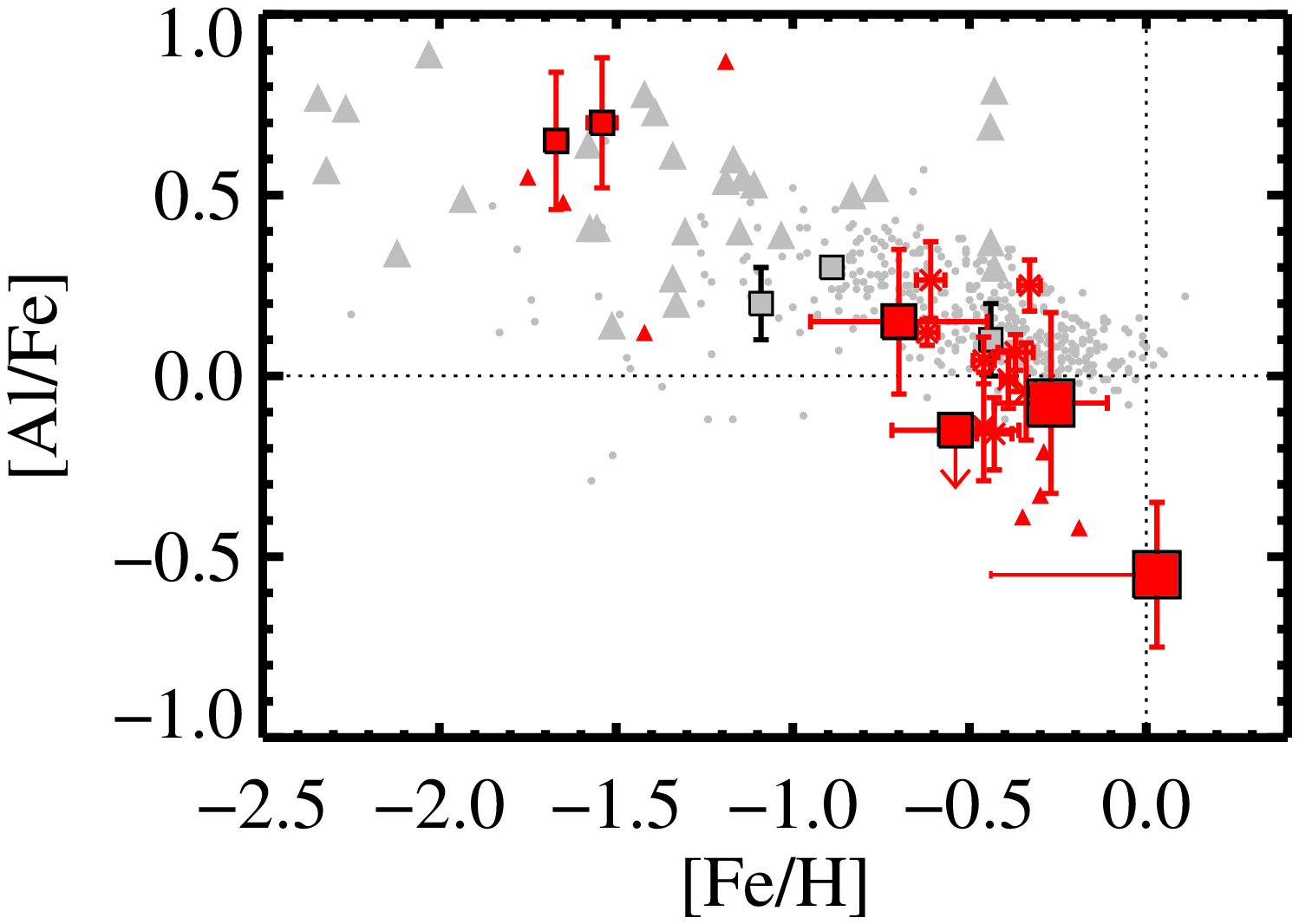}
\caption{\footnotesize Comparison of detailed abundances in the MW,
  M31, and the LMC.  Stellar MW abundances from \cite{venn} are shown
  as small gray circles, stellar GC abundances from \cite{pritzl} are
  shown as gray triangles, and MW IL abundances from \cite{5} are
  shown as gray squares. In left panels, M31 GC IL abundances are
  shown as green squares. In right panels, LMC GC abundances are shown
  as red squares, with increasing symbol size denoting younger
  ages. The total range in age is 0.05$-$12 Gyr. LMC stellar
  abundances from \cite{pompeia}, \cite{mucc1,mucc2,mucc3} and
  \cite{jj06} are shown as small red symbols.}
\label{fig:abund}
\end{figure*}

\vspace{-0.1cm}
In the right panels of Figure \ref{fig:abund}, we show our LMC IL
abundances from \cite{7}. The [Ca/Fe] and [Si/Fe] of the LMC decrease as a function
of [Fe/H] more than is seen in M31 or the MW.  This suggests 
a slower, more prolonged star formation history in the LMC than in 
more massive galaxies.  In addition, the LMC sample contains clusters
with much younger ages. 

The oldest GCs in the LMC show [Na/Fe] and [Al/Fe] that are
elevated to similar levels as seen in the MW and M31 GCs. However, the
younger LMC GCs, which also happen to be the least massive GCs, show
much lower values of [Na/Fe] and [Al/Fe]. This may provide insight
into the formation scenario that results in abundance variations,
however it is unclear if age or total GC mass is the dominant factor.
Detailed abundances of young massive GCs will provide a crucial next
test.

\begin{figure*}
\vskip -0.1cm
\includegraphics[clip=true,scale=0.35]{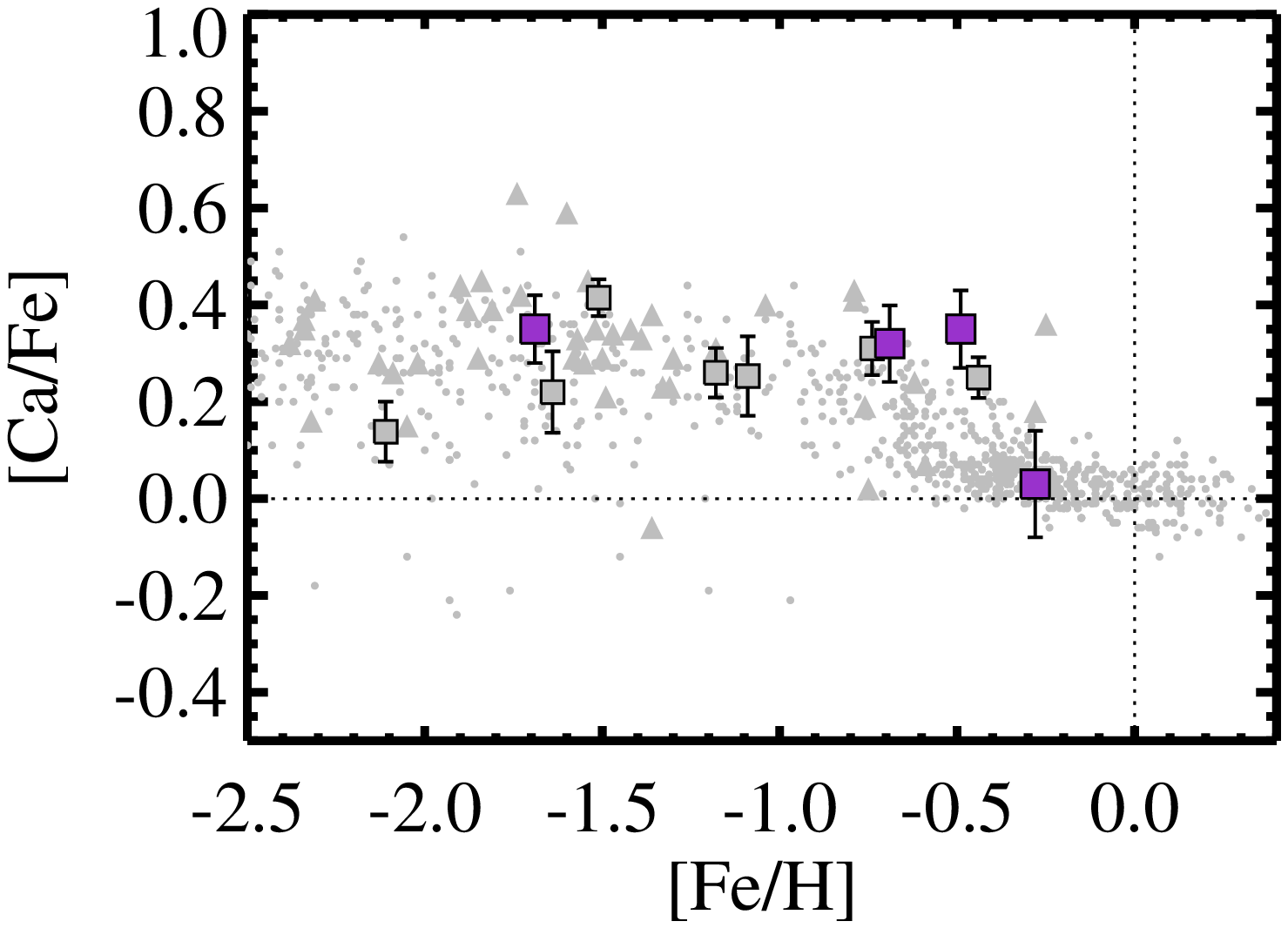}
\includegraphics[clip=true,scale=0.35]{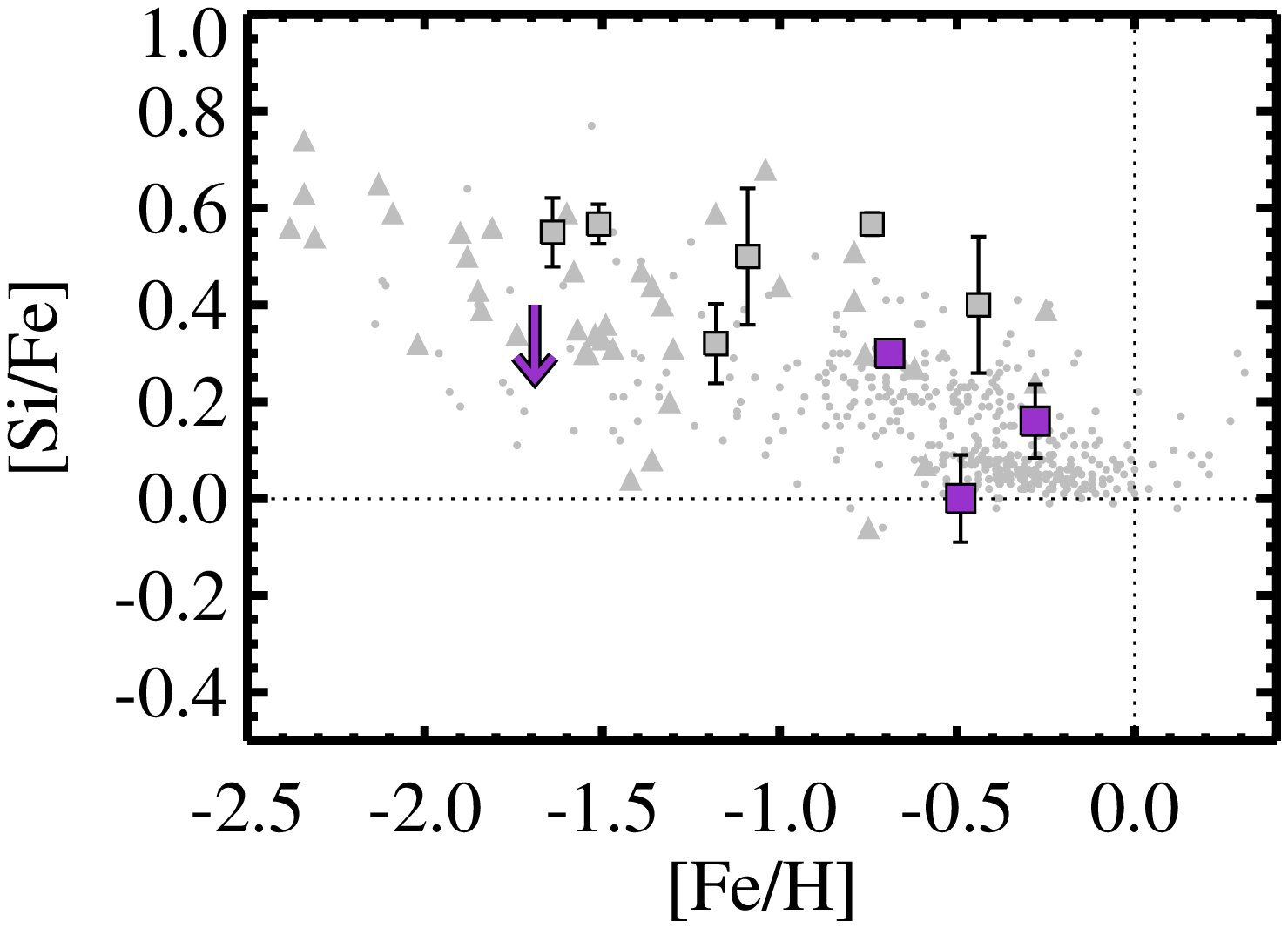}
\vskip -0.3cm
\caption{\footnotesize Purple squares show preliminary Ca and Si abundances in NGC5128. Gray symbols are as in 
  Fig 1.}
\label{fig:5128}
\end{figure*}

\subsection{NGC 5128 (Massive E/S0)}
In Figure \ref{fig:5128}, we show preliminary abundances of Fe, Ca,
and Si for 4 GCs in NGC 5128.  These are the first $\alpha$-element
abundances derived from individual Fe, Ca, and Si lines in a massive
E/S0 type galaxy.  This is a subset of the GCs currently in our sample
from an ongoing program to study NGC 5128, the first 10 GCs of
which will be presented in Colucci et al. (2013). Our preliminary
results indicate that [$\alpha$/Fe] in the more metal-poor GCs is
comparable to that in the MW and M31 GCs, and that [$\alpha$/Fe] may
decrease at higher [Fe/H].  This suggests that the star formation rate
in NGC 5128 at early times was was comparable to that of the MW and
M31.

 \section{Summary}
 \vspace{-0.1cm}
 Detailed chemical abundances from high resolution IL spectra of GCs
 can be obtained in galaxies with a range of masses and Hubble
 types. This makes it possible for quantitative comparisons of
 detailed chemical enrichment histories and cluster formation
 scenarios in galaxies of all Hubble types and environments.
 
\begin{acknowledgements}
  We thank the organizing committee and INAF for the opportunity to
  present this work and for hosting a very interesting conference.
\end{acknowledgements}

\bibliographystyle{aa}

\end{document}